\documentclass[%
 letterpaper,
superscriptaddress,
amsmath,amssymb, aps,nobalancelastpage,twocolumn,
pra,
]{revtex4-2}

\usepackage{braket}
\usepackage{graphicx}
\usepackage{dcolumn}
\usepackage{bm}

\usepackage{adjustbox}
\usepackage{hyperref}
\usepackage{subfigure}
\usepackage{todonotes}
\usepackage{setspace}

\usepackage{soul}
\usepackage[normalem]{ulem}

\begin{document}

\title{Benchmarking memory and nonlinear processing capacity of quantum recurrent neural networks for time series prediction}

\newcommand{\cesga}{Galicia Supercomputing Center (CESGA), 15705 Santiago de Compostela, Spain}
\newcommand{\cograde}{Computer Graphics and Data Engineering (COGRADE), Departamento de Electrónica e Computación, Universidade de Santiago de Compostela, 15782 Santiago de Compostela, Spain}

\author{J.D. Viqueira}
\email{josedaniel.viqueira@rai.usc.es}
\affiliation{\cesga}
\affiliation{\cograde}
\author{A. G\'omez }
\affiliation{\cesga}
\author{D. Mera}
\affiliation{\cograde}

\date{\today}

\begin{abstract}
Quantum machine learning has emerged as a promising approach to time series prediction. While quantum reservoir computing has been widely explored for this setting, variational quantum models tailored to sequential data remain comparatively understudied, largely because they are difficult to train. In this work, we present a rigorous benchmark of variational quantum recurrent neural networks (QRNNs) on time series prediction tasks, focusing on two key capabilities: memory and nonlinear processing. We introduce a task-driven methodology that quantifies model capacity, enabling an intuitive and fair comparison across architectures, including classical recurrent baselines. We evaluate the proposed framework on two generic benchmark tasks and on a synthetic dataset describing a chaotic system. Overall, the quantum models achieve performance comparable to their classical counterparts, but classical models scale better with the memory size within the range studied. Notably, one QRNN variant shows higher average capacity on the task designed to jointly probe memory and nonlinear capacity. Moreover, our results show that quantum models offer better capacity when we are able to find a good parameter initialization. However, the increasing sensitivity to initialization leads to a lower probability of finding good solutions as the model size increases. These results provide a principled starting point for designing richer quantum recurrent architectures and for assessing where they may offer practical benefits.
\end{abstract}

\maketitle


\section{Introduction \label{sec:intro}}

Time series processing has been a central problem in machine learning because of its complexity and the dynamic nature of real-world data. Autoregressive integrated moving average (ARIMA), long short-term memory (LSTM) networks, and Transformers are very different methods, but all have been successful for numerical time series forecasting and text generation in different contexts \cite{zhang_time_2003,hochreiter_long_1997,vaswani_attention_2017}.

Despite these successes, there is still room to improve performance and efficiency. Moreover, improved performance often comes with greater architectural complexity, reducing interpretability and increasing the computational cost of architecture search. Quantum computing is an alternative paradigm that may enable new efficient algorithms, either developed from scratch or inspired by simpler models \cite{schuld_supervised_2018}.

Most work on time series forecasting with quantum computing has focused on quantum reservoir computing (QRC) \cite{fujii_harnessing_2017}. QRC greatly simplifies training by reducing it to a convex optimization problem. A (fixed) quantum substrate generates high-dimensional feature vectors with nonlinear dependence on the input time series, and a trained linear readout layer maps these vectors to low-dimensional outputs. Research on QRC ranges from theoretical foundations \cite{grigoryeva_echo_2018,sannia_dissipation_2024,sannia_non-markovianity_2026,monzani_universality_2025,sannia_exponential_2025,xiong_role_2025} to numerical benchmarks \cite{mujal_time-series_2023,kobayashi_feedback-driven_2024,garcia-beni_squeezing_2024}. Benchmarks typically rely on classical and quantum chaotic systems, as well as synthetic time series generated by auto-regressive moving-average (ARMA) processes \cite{brockwell_introduction_2016}.

In contrast, variational quantum recurrent neural networks (QRNNs) have mainly been studied in specific use cases and benchmarked for time series prediction \cite{bausch_recurrent_2020, takaki_learning_2021, chen_quantum_2022, li_quantum_2023, viqueira_density_2025, fellner_quantum_2026}. However, intermediate-measurement-based variational QRNNs with two qubit registers --i.e., two subsets of qubits, one acting as memory and the other used for input encoding and reset-based measurement-- remain comparatively less understood \cite{viqueira_density_2025}. Hereafter, the term QRNN refers to this specific model family. New benchmarks are therefore needed to characterize how performance scales with task difficulty and system size, and to compare fairly against classical baselines. In particular, QRC benchmarks \cite{mujal_time-series_2023,kobayashi_feedback-driven_2024,garcia-beni_squeezing_2024} provide methodology that can be adapted to variational recurrent models, which share the recurrent structure but differ in their training strategy.

Both memory and nonlinear processing capacity of models used for forecasting are of crucial relevance, because time series generated in complex and real-world systems are often governed by complex patterns. These patterns are correlated with the past, and they need a basis of nonlinear functions to be described mathematically. A model with high-performance memory can retain accurate dependencies on past inputs, while it can reproduce the complex patterns with a high capacity of generating nonlinear functions. The main aim of this article is to provide a rigorous study of the QRNN model family and to assess its memory and nonlinear processing capacities for time series forecasting. To this end, we develop a methodology based on multiple experiments that test different models and configurations across several datasets with a tunable level of temporal dependence. Relating performance to task difficulty allows us to quantify the desired properties of each model and configuration.

The manuscript is structured as follows. In Sec. \ref{sec:background} we introduce the QRNN, discuss vanishing gradients as a potential cause of poor training behavior, and define quantum mutual information through time, which will be used to manage datasets before training. In Sec. \ref{sec:methods} we define the tasks, the models under study, and the methodology. Sec. \ref{sec:results} summarizes the results and compares the quantum models with well-known classical baselines. Finally, Sec. \ref{sec:conclusions} summarizes the conclusions and future work.

\section{Quantum recurrent neural network: limitations and potential advantages \label{sec:background}}

Recurrent models are intended to be used with time-ordered data, i.e., time series, composed of $n_T$ timesteps, each labeled with $t$. They are defined as discrete sets
\begin{equation}
    X = \{ \bm{x}_{(t)} \}_{t=1}^{n_T} \; | \; \bm{x}_{(t)} \in \mathbb{R}^{d_x}.
\end{equation}

Recurrent models consist of functions that, given an input and a history (a vector containing information about previous inputs), generate a new value. To make these functions trainable, we design them as neural networks containing a set of trainable parameters $\bm{\theta}$ that are iteratively changed when training the model. In classical neural networks they are commonly known as weights.

In general, when inputs are multivariate and outputs are univariate, a classical recurrent neural network (CRNN) implements a function
\begin{equation}
\begin{array}{lll}
    f_{\bm{\theta}} &: \mathbb{R}^{d_x} \times \mathbb{R}^{d_h} & \to \mathbb{R} \times \mathbb{R}^{d_h} \\
    & \left( \bm{x}_{(t)}, \bm{h}_{(t-1)} \right) & \mapsto \left( y_{(t)}, \bm{h}_{(t)} \right) ,
\end{array}
\end{equation}
applied  recursively every timestep. $\bm{x}_{(t)}$ and $y_{(t)}$ are the input vector and output value at timestep $t$, while $\bm{h}_{(t)}$ is a memory vector (hidden state) of dimension $d_h$. This memory vector can be a concatenation of several vectors undergoing independent processes, as occurring in LSTM and gated recurrent unit (GRU) models \cite{cho_learning_2014}.

Meanwhile, a QRNN applies the function
\begin{equation}
\begin{array}{lll}
    g_{\bm{\theta}} &: \mathbb{R}^{d_x} \times L(\mathcal{H}_{M}) & \to \mathbb{R} \times L(\mathcal{H}_{M}) \\
    & \left( \bm{x}_{(t)}, \rho_{(t-1)} \right) & \mapsto \left( y_{(t)}, \rho_{(t)} \right) 
\end{array}
\end{equation}
recursively at every timestep. Here, $\rho_{(t)}$ is a density matrix in $L(\mathcal{H}_{M})$, the space of bounded linear operators on the memory Hilbert space $\mathcal{H}_{M}$ of dimension $N_M = 2^{n_M}$. The parameter $n_M$ denotes the number of memory qubits.

\begin{figure*}
    \centering
    \includegraphics[width=\linewidth]{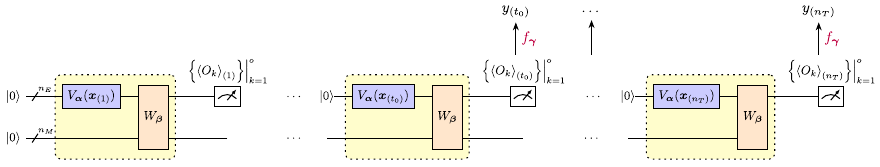}
    \caption{QRNN diagram. Blue, orange, and purple denote the configurable encoding, evolution, and output components, respectively. The last $n_p = n_T - t_0 + 1$ timesteps form the prediction window.}
    \label{fig:qrnn_circuit}
\end{figure*}

The core of the QRNN is a recurrent quantum circuit with two registers: exchange (E) and memory (M). The circuit is constructed as follows (Fig.~\ref{fig:qrnn_circuit}):
\begin{enumerate}
    \item Initialize register M in the state $\ket{0}^{\otimes n_M}$.
    \item Reset register E to the state $\ket{0}^{\otimes n_E}$.
    \item Apply a parameterized unitary (ansatz) $U_{\bm{\theta}'}\!\left(\bm{x}_{(t)}\right)$ that evolves the qubits and entangles registers E and M, being $\bm{\theta}'$ a subset of the complete set of trainable parameters $\bm{\theta}$, and $t$ the current timestep.
    \item Measure the qubits in register E to obtain a set of observables.
    \item Repeat steps 2--4 for $t=2,\dots,n_T$.
\end{enumerate}

In our benchmark, the QRNN family is defined as a quantum-circuit-based model with three configurable components: (1) an encoding operator, (2) an evolution operator, and (3) an output layer. (1) and (2) are unitary operators and together form the full ansatz $U_{\bm{\theta}'}\!\left(\bm{x}_{(t)}\right)$. The encoding operator prepares a pure state in register E that depends on the input,
\begin{equation}
    V_{\bm{\alpha}} (\bm{x}_{(t)}) \in L(\mathcal{H}_E) , \; \; \bm{\alpha} \subset \bm{\theta} ,
\end{equation}
while the evolution operator entangles both quantum registers and evolves the full system state,
\begin{equation}
    W_{\bm{\beta}} \in L(\mathcal{H}_{EM}) , \; \; \bm{\beta} \subset \bm{\theta},
\end{equation}
where $\mathcal{H}_{EM} = \mathcal{H}_E \otimes \mathcal{H}_M$.

The output layer is a linear map from a real feature vector to a scalar,
\begin{equation}
\begin{array}{lll}
     f_{\bm{\gamma}} &: \mathbb{R}^o & \to \mathbb{R} \\
     & \left(\braket{O_1}, \dots, \braket{O_o} \right) & \mapsto \gamma_0 + \sum_k \gamma_k \braket{O_k} ,
     \label{eq:output_layer_general}
\end{array}
\end{equation}
where $\gamma_k \in \bm{\gamma} \subset \bm{\theta}$ are the remaining trainable parameters and $\braket{O_k}$ are expectation values of observables $O_k \in L(\mathcal{H}_E)$. In general, estimating multiple expectation values requires running a set of circuits with measurements in different bases.

Recurrent models for time series prediction require a dissipative memory so that outputs can depend (strongly or weakly) on past inputs. Typically, recent events have a larger influence on the present, but long-term dependencies can also occur. Therefore, models need a finite-memory mechanism that gradually discards irrelevant information while retaining salient features.

When training with gradient-based methods, long-term dependencies are difficult to learn because of the vanishing-gradient problem, which has been extensively studied in CRNNs \cite{hochreiter_long_1997,bengio_learning_1994}.

In quantum machine learning, there is extensive work on QRC, where memory capacities are assessed numerically \cite{fujii_harnessing_2017,mujal_time-series_2023,kobayashi_feedback-driven_2024,garcia-beni_squeezing_2024}. There is also substantial literature on barren plateaus in variational quantum algorithms (VQAs) \cite{mcclean_barren_2018,cerezo_higher_2021,pesah_absence_2021,arrasmith_effect_2021,sannia_engineered_2024}. However, there is comparatively less work on vanishing gradients and memory dissipation in variational quantum recurrent structures \cite{li_quantum_2025,nikoloska_time-warping_2023}.

In this section, we first summarize the vanishing-gradient problem in recurrent neural networks to motivate benchmarking QRNNs against classical baselines. We then introduce quantum mutual information (QMI) through time, by analogy with its use in quantum communication channels. QMI is crucial for choosing the sample length and the number of predicted timesteps per sample: the recurrent structure (the QRNN circuit) must match the sample length. Physical quantities such as QMI therefore provide a principled way to split a dataset into fixed-length samples.

\subsection{The memory challenge \label{subsec:vanish_grads}}

The difficulty classical neural networks have in learning long-term dependencies is commonly attributed to gradient evaluation via backpropagation through time (BPTT) \cite{hochreiter_long_1997}. Vanishing or exploding gradients arise from repeated multiplication by weight matrices $W$ \cite{zhang_dive_2023}: eigenvalues larger than 1 lead to divergence, while eigenvalues smaller than 1 lead to vanishing derivatives.

For bounded observables and the parameterized circuit family considered here, the recurrent update does not by itself imply the same unbounded gradient growth associated with repeated multiplication by arbitrary classical weight matrices. Vanishing gradients, however, can still occur. In VQAs, they are well known, either due to local minima or barren plateaus \cite{failde_using_2023, mcclean_barren_2018}.

Analytical gradient evaluation in QRNNs can be formulated as a generalization of the parameter-shift rule \cite{schuld_evaluating_2019}. This provides a quantum analogue of BPTT, in which the chain rule is applied recursively to quantum operators.
It was shown that numerical gradients computed via finite differences match analytical gradients in complete training loops when evaluating ideal circuits in simulation. Under shot noise, results are comparable and improve as the number of shots increases \cite{viqueira_density_2025}.

Intuitively, as time progresses, gradients may vanish and dependence on earlier inputs may decrease because the system loses information. The extent of this effect depends strongly on the network structure (i.e., the internal operator design) and on the particular task.

A central question in this article is how memory size affects model performance. Reproducing an input from $\tau$ timesteps in the past --the short-term memory task defined later-- is trivial if the model uses an explicit memory vector of size $d_m \geq \tau$, since the memory can act as a buffer that stores the last $\tau$ inputs. A CRNN therefore requires memory that scales linearly with $d_m$. A natural question is whether a quantum model can implement such memory with only $\log d_m$ qubits. Quantum states live in an exponentially large Hilbert space and therefore have exponentially many degrees of freedom that could, in principle, be exploited by sufficiently expressive circuits. However, vanishing gradients can severely limit trainability. Our benchmarks estimate how likely it is to find recurrent configurations that successfully exploit these degrees of freedom.

\subsection{Mutual information in QRNN}

Periodic monitoring of the quantum circuit through projective measurements causes its dynamics to escape from unitary evolution. Despite non-unitary operations having harmful effects in quantum computing due to the loss of information, it may be beneficial for models that process time series. In the area of QRC, it has been shown that open quantum systems offer desirable properties for having good reservoirs \cite{sannia_dissipation_2024, ricci_quantum_2026}. In contrast to reservoir computing, in QRNNs, the training process changes the circuit dynamics itself.

A good measure of the statistical uncertainty within the description of a quantum system is the von Neumann entropy, a generalization of classical entropy to quantum states. For a quantum state described by a density matrix $\rho$, it is defined by
\begin{equation}
    S(\rho) \equiv - \mathrm{tr} \left( \rho \log \rho \right) .
\end{equation}

We are working with a bipartite system, where a quantum register, E, is periodically measured and the other one, M, is subject to unitary evolution while interacting with the first one. System E could indeed be though of as an environment that makes M leave unitary dynamics. However, in contrast to common open systems where the main system is what matters, here the system E is as important as M, because information contained in M can only be accessed indirectly through E. In summary, a single measurement of entropy is not enough to describe what is happening in the whole system. This is why we use the mutual information (MI). In classical information and probability theory, given two random variables $X$ and $Y$, MI measures how much information do $X$ and $Y$ have in common. Similarly, quantum MI (QMI) (also von Neumann MI) measures the correlation between two subsystems $A$ and $B$ of a quantum state. It is defined by
\begin{equation}
    S(A:B) \equiv S(A) + S(B) - S(A, B),
\end{equation}
where $S(A,B)$ is the entropy of the whole system.

In simulation, we monitor the QMI in every timestep of the QRNN just before measurement. This allows us to quantify the correlation between exchange and memory registers before resetting the first one to encode the data for the next timestep.

At timestep 1, the entropy of the whole system is null, $S(E,M)=0$, because the state is pure. It has not undergone any dissipative operation. Provided that both registers are entangled, the entropies of both registers are nonzero, $S(E), S(M) > 0$. After timestep 1, the system will no longer be in a pure state, then $S(E,M) > 0$. Apart from that, all the entropies are upper-bounded by $\log d$, being $d$ the dimension of the system. In addition, the QMI satisfies \cite{wilde_classical_2011}

\begin{equation}
\begin{split}
         0 \leq S(E:M) \leq \\ 2 \min \left( \log \left( \dim (\mathcal{H}_E) \right) , \log \left( \dim (\mathcal{H}_M) \right) \right) .
\end{split}
\end{equation}

The behavior strongly depends on the system structure. In the case of a QRNN, which has a fixed structure but whose dynamics depend on dozens or thousands of tunable parameters, it is difficult to predict. However, following the theory of open quantum systems, we know that an open quantum system (e.g., a system interacting with a thermal bath) converges to a stationary state \cite{breuer_theory_2002,overbeck_time_2016}. A QRNN is expected to exhibit similar behavior due to its interaction with the classical world through projective measurements.

One of the desirable properties of a reservoir system for time series processing is the echo-state property, which implies that the system does not depend on the initial conditions of the reservoir after a relaxation or washout time. Similarly, the QRNN system transits from a state initially pure to a highly mixed state, losing the effects of the initial state. The QMI will allow us to determine when the stationary phase has begun and will mark the point when we can start to begin evaluating predictions.

\section{Methodology \label{sec:methods}}

We first describe the general methodology and then specify the experimental settings that produced our results in Sec.~\ref{sec:results}.

We define a comprehensive set of experiments consisting of 4 QRNN models and 3 well-known classical models on three univariate time-series tasks. We test several configurations (hyperparameter sets) per model, limited by the amount of time a set of experiments needs to be computed. Each experiment is performed on a classical machine, simulating either a classical or a quantum recurrent neural network. Details of the software and hardware are provided in App.~\ref{app:simulation_details}.

We start the methodology by \textit{data preparation}, which includes: (i) synthetic data generation or real-world data gathering, (ii) data pre-processing with normalization, and (iii) splitting the data into training, validation and test sets.

When splitting between training, validation and test sets, we make sure that the smallest set is large enough so that the statistical fluctuations of the cost and metrics are below some threshold, assuming that every data point is independent and identically distributed (i.i.d.).  $\mathcal{S}_\text{tra}$, $\mathcal{S}_\text{val}$, $\mathcal{S}_\text{tst}$ denote the collections of indices identifying the samples from the original dataset contained in each set.

After splitting, the datasets are structured collections of disjoint time series samples, each represented by a tuple
\begin{equation}
(X_s, Y_s) \: 
\left\{
\begin{array}{lll}
     X_s &= \{ x_{s (t)} \}_{t=1}^{n_T} \; | \; x_{s (t)} \in [-1,1] & \; \text{(input)} \\
     Y_s&= \{ y_{s (t)} \}_{t=1}^{n_T} \; | \; y_{s (t)} \in [-1,1] & \; \text{(target)} ,
\end{array}
\right.
\end{equation}
where $x_{s (t)}$ are no longer vectors, because we restrict the analysis to univariate series (the input per timestep is a scalar).

In the \textit{training phase}, every time series sample from the collection $\{ \left( X_s, Y_s \right) \}_{s \in \mathcal{S}_\text{tra}}$ is processed by a single recurrent instance at once. The recurrent structure sequentially processes an input value and returns a value per timestep $t$. Thus, the dimensions of input and output layers must agree with the time series sample length.

The training routine aims to minimize a cost function that quantifies an overall distance between the predictions and the targets. To this end, the routine iterates over every sample in $\mathcal{S}_\text{tra}$ computing the gradient of a loss function. The loss function measures a distance between the predictions $\bar{Y}_s$ and the targets $Y_s$ considering the last $n_p$ timesteps,
\begin{equation}
    L \equiv \text{dist} \left( \bar{Y}_s', Y_s' \right) ,
\end{equation}
where the prime means that the sample was reduced to the last $n_p$ timesteps. That is, a reduced sample is defined as
\begin{equation}
    Y_s' \equiv \{ y_{s (t)} \in Y_s \}_{t=t_0}^{n_T},
\end{equation}
where $t_0 \equiv n_T-n_p+1$. The reason for considering only the last $n_p$ points is that the recurrent network needs to be in a stationary state and needs past context for data.  Predictions at timesteps $t < t_0$ are excluded to compute the loss to allow for relaxation after an initial transitory phase of the QRNN circuit and to provide temporal context.

The training ends after running a fixed number of epochs $N_\text{ep}$. An epoch ends when the gradient has been evaluated over the $N_\text{tr}$ samples from the training set in random order (shuffling is enabled) and the cost function is evaluated as a distance between all the predictions and targets in both the training and the validation sets, $\mathcal{S}_\text{tra}$ and $\mathcal{S}_\text{val}$, respectively,
\begin{equation}
\begin{array}{ll}
     C_\text{tra} &= \text{dist} \left( \bigcup_{s \in \mathcal{S}_\text{tra}} \bar{Y}_s', \bigcup_{s \in \mathcal{S}_\text{tra}} Y_s' \right) \\
     C_\text{val} &= \text{dist} \left( \bigcup_{s \in \mathcal{S}_\text{val}} \bar{Y}_s', \bigcup_{s \in \mathcal{S}_\text{val}} Y_s' \right)
\end{array} .
\end{equation}

To avoid overfitting, instead of declaring the set of parameters after the last epoch as the solution, we choose the epoch where $C_\text{val}$ is minimum, thus finding
\begin{equation}
    \min_{\bm{\theta}} C_\text{val}.
\end{equation}

Every single training is part of a battery of training runs, where different recurrent models, with different hyperparameters and different datasets are compared through a metric $M$ evaluated on the test set $\mathcal{S}_\text{tst}$, denoted as $M_\text{tst}$. Moreover, every single training depends on random components, which are random initialization of the trainable parameters and sample shuffling. This is why we repeat the experiments $N_\text{exp}$ times.

These sets of training runs allow us to quantify memory, nonlinear processing and prediction capacities as performance scores on three different tasks, as well as QMI for each model. The number of experiments is summarized in Tab.~\ref{tab:all_experiments}.

\begin{table*}[]
\small
\begin{tabular}{|l|c|c|c|r|}
\hline
\textbf{Model name} & STM \cite{mujal_time-series_2023}   & NARMA \cite{nakajima_boosting_2019}   & MG \cite{garcia-beni_squeezing_2024}    & \textbf{Subtotal}  \\ \hline \hline
EMCZ3           & $6 \times 11 \times 10$     & $6 \times 10 \times 10$     & $6 \times 11 \times 10$                 & 1,920      \\ \hline
EMCZ2 \cite{viqueira_density_2025}        & $6 \times 11 \times 10$     & $6 \times 10 \times 10$     & $6 \times 11 \times 10$                 & 1,920      \\ \hline
EMCX1                                       & $6 \times 11 \times 10$     & $6 \times 10 \times 10$     & $6 \times 11 \times 10$                 & 1,920      \\ \hline 
Hybrid & $6 \times 11 \times 10$ & $6 \times 10 \times 10$ & $6 \times 11 \times 10$ & 1,920 \\ \hline \hline
VanillaRNN \cite{elman_finding_1990}        & $6 \times 11 \times 10$      & $6 \times 10 \times 10$      & $6  \times 11 \times 10$                 & 1,920      \\ \hline
LSTM \cite{hochreiter_long_1997}            & $6 \times 11 \times 10$      & $6 \times 10 \times 10$      & $6  \times 11 \times 10$                 & 1,920      \\ \hline
GRU \cite{cho_learning_2014}                & $6 \times 11 \times 10$      & $6 \times 10 \times 10$      & $6 \times 11 \times 10$                 & 1,920      \\ \hline \hline
\textbf{Subtotal}                                    & 4,620                        & 4,200                        & 4,620                                    & \textbf{13,440} \\
 \hline
\end{tabular}

\caption{Summary of models and datasets for benchmark. Each cell contains (number of model variants) $\times$ (number of dataset variants) $\times$ (number of repetitions for statistics, $N_\text{exp}$).}
\label{tab:all_experiments}
\end{table*}

We set the number of timesteps per sample, $n_T$, and the number of output timesteps included in the loss function, $n_p$, according to two criteria: (i) the $n_p$ output timesteps should not depend on the recurrent network's initial conditions, and (ii) they should have enough context from the data.

First, we define the transitory phase as the number of timesteps during which a random network remains in a transient state before reaching a stationary one. A quantum random recurrent model (QRRM) is a quantum recurrent model in which the encoding operator $V$ is the identity on register E and the evolution operator $W$ is a Haar-random unitary on the full space sampled through the QR-decomposition method \cite{mezzadri_how_2007}. The set of parameters $\bm{\gamma}$ and the set of observables do not affect entropy measurements and can be set as empty.
Because of its structure, the network interacts with the environment at every timestep through projective measurements. Although it starts in a pure state $\ket{0}^{\otimes n_E} \otimes \ket{0}^{\otimes n_M}$, the entropy of the quantum system increases as intermediate measurements occur. In the absence of structure, on average, the entropies of subsystems E, M, and EM increase until they saturate. Consequently, the QMI also saturates to a value that is not zero (its theoretical lower bound). Once these limits are reached, the transitory phase has been exceeded.

Second, the required context depends on the dataset. ARMA processes --the first two datasets used here belong to this family-- are defined by a parameter $\tau$ that strictly bounds the number of timesteps over which inputs and outputs are correlated. In other datasets, $\tau$ may correspond to a delay in the governing differential equation or to the prediction horizon. In general, $\tau$ controls the difficulty of the prediction task.

Finally, we \textit{test and quantify the capacity} of the models to solve a task. For a given model, each hyperparameter set defines an architecture that is trained $N_\text{exp}$ times, each time starting from a different random parameter initialization. This initialization leads to variability in the final results. The benchmarks are designed to study how performance depends on memory size, defined as the number of qubits (quantum models) or the number of hidden-state vector components (classical models). Following Sec.~\ref{subsec:vanish_grads}, in the STM task we expect maximization metrics (such as $R^2$) to saturate for $\tau \leq \tau_\text{max}$ and then, on average, decay for $\tau > \tau_\text{max}$. The average metric behavior is described by a continuous function $\mu(\tau)$ for $\tau \in \mathbb{R}_{\geq 0}$. For each $\tau$, the observed values follow a distribution, and $\mu(\tau)$ should reproduce the mean of these distributions at each point.

The \textit{capacity} of a model is defined as the $\tau$ at which $\mu (\tau)$ decays below a given threshold, $\varphi_c$. This $\tau$ is understood as a cutoff,
\begin{equation}
    \tau_c = \min \{ \tau \in \mathbb{R}_{\geq 0} \: | \: \mu(\tau) <  \varphi_c \} .
\end{equation}
Because the fitted curve need not be strictly monotonic, this is a threshold-crossing measure rather than a guarantee of monotonic degradation.

In addition, due to the dispersion of the metrics distributions for each given $\tau$, we also define the cutoffs of their percentiles analogously.
We repeat the methodology for the performance against the number of trainable parameters to provide an approximate comparison of the gradient computation cost.

\subsection{Data preparation}

We define two tasks to evaluate and compare different model properties. Each task consists of training the models on a family of datasets parameterized by a generation parameter $\tau$.

The first task, \textit{short-term memory} (STM), serves as an isolated benchmark for memory \cite{mujal_time-series_2023}. The goal is to reproduce the input time series delayed by $\tau$ timesteps. The STM-$\tau$ series is generated by
\begin{equation}
    y_{(t)} = x_{(t-\tau)} \; | \; x_{(t)} \sim \mathcal{U} (-1,1) .
\end{equation}

The second task, \textit{nonlinear autoregressive moving average} (NARMA), serves as a benchmark for memory and nonlinear processing capacity. The NARMA-$\tau$ series is generated by
\begin{equation}
\begin{split}
     x_{(t)} &\sim \mathcal{U} (-1,1) , \; u_{(t)} = 0.04 \: x_{(t)}, \\
     y_{(t)} &= \alpha_\text{NA} y_{(t-1)} + \\
     &+ \beta_\text{NA} y_{(t-1)} \sum_{j=-\tau}^0 y_{(t+j)} + \gamma_\text{NA} u_{(t-1)} u_{(t-\tau)} + \delta_\text{NA} ,  
\end{split}
\end{equation}
where the parameters are
\begin{equation}
    \alpha_\text{NA} = 0.3, \; \beta_\text{NA} = 0.05, \; \gamma_\text{NA} = 1.5, \; \delta_\text{NA} = 0.1 ,
\end{equation}
which ensures controlled stability, long-term dependencies, and high nonlinearity, following standard choices in the literature \cite{atiya_new_2000,fujii_harnessing_2017,garcia-beni_squeezing_2024}.

For both datasets, we study the range $\tau \leq 10$. We omit $\tau=0$ in the NARMA task because it is equivalent to STM-0 with an offset.

In addition to these two tasks, we benchmark the models on a third dataset: the \textit{Mackey--Glass} (MG) time series, a common benchmark for forecasting models \cite{garcia-beni_squeezing_2024}. It is generated by a delay differential equation in which the delay $\tau_\text{MG}$ determines the dynamical regime. The system becomes chaotic for $\tau_\text{MG} \approx 17$, which is the value we fix in our experiments. To increase task difficulty, we vary the prediction horizon $\tau$, i.e., the delay of the targets with respect to the inputs, and we define the task as predicting
\begin{equation}
    y_{(t)} = x_{(t+\tau)} .
\end{equation}
The MG series is a discretization of the continuous process governed by the differential equation
\begin{equation}
    \frac{dx}{dt} = \frac{\beta_\text{MG} x(t-\tau_\text{MG})}{1 + x^{n_\text{MG}}(t-\tau_\text{MG})} - \gamma_\text{MG} x(t)
\end{equation}
with parameters
\begin{equation}
    \tau_\text{MG} = 17, \; n_\text{MG}=10, \; \beta_\text{MG}=0.2, \; \gamma_\text{MG}=0.1 ,
\end{equation}
and initial condition of $x(t \leq 0)=1.1$. Under this regime, the Lyapunov time is $140 \leq 1/\lambda_\text{max} \leq 160$ in natural units; it provides a theoretical prediction horizon beyond which nearby trajectories diverge by a factor $e$. With a sampling rate of 5.33 for the MG signal, the Lyapunov time is reached after approximately 30 timesteps. For this reason, we extend the range of $\tau$ up to 60 (twice the Lyapunov time). We use a step size of 6, yielding 11 values of $\tau$ to obtain metrics as a function of $\tau$.

Both NARMA and MG are Z-score normalized. We then discard the first 20 timesteps in NARMA and the first 3000 timesteps in MG to ensure stationarity. Finally, we rescale all three datasets so that the maximum absolute value is 0.95. This avoids predictions extremely close to $\pm 1$, which are strict bounds when evaluating Pauli expectation values in QRNNs.

For each task, we generate 3000-timestep series. In STM and NARMA, the input series are i.i.d. We split the data into training, validation, and test sets with proportions 70\%, 15\%, and 15\%, respectively. The test set therefore contains $n_\text{tst}=450$ timesteps. For STM and NARMA we use $\tau \leq 10$; in STM we also evaluate $\tau=0$, corresponding to the trivial identity mapping.

\begin{figure}
    \centering
    \includegraphics[scale=1]{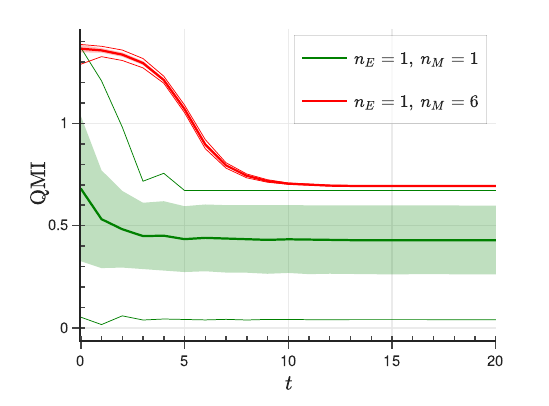}
    \caption{QMI for representative sizes of random recurrent networks. Thick lines show the mean over 100 experiments; shaded bands indicate one standard deviation; thin lines show the minimum and maximum values.}
    \label{fig:max_washout}
\end{figure}

We compare quantum models with $2 \leq n_E + n_M \leq 7$, motivated by simulation and training time constraints (maximum total number of qubits: 7). We estimate the maximum transitory phase across all model sizes by evaluating the QMI as a function of timestep in a QRRM.
As representative examples, Fig.~\ref{fig:max_washout} shows QMI versus timestep for the minimum and maximum transitory times for $n_M \in \{1,6\}$.

In the STM and NARMA tasks, outputs depend on inputs up to lag $\tau$, so the maximum context length required is 10 timesteps. For MG, this value is not known a priori; here we test model prediction capacity given a fixed QRNN length $n_T$.

Based on these considerations, we set 10 timesteps for the transitory phase, 10 timesteps for context, and a prediction window of $n_p=10$ output timesteps per sample. Therefore, $n_T=30$. We start the context window after the transitory because input dependencies can be affected by transient dynamics. For a fair comparison, we use the same $n_T$ and $n_p$ in the CRNN baselines.

Thus, the number of output points (timesteps) used to compute cost and metrics is
\begin{equation}
    N_\text{tst} n_p = 450 / 30 \times 10 = 150 .
\end{equation}

We use the root mean square error (RMSE) as the cost. Assuming predictions and targets are i.i.d. and uniformly distributed (the worst case, corresponding to an untrained model), the RMSE is approximately normally distributed by the central limit theorem,
\begin{equation}
    \text{RMSE} \sim \mathcal{N} \left( \sqrt{\frac{2}{3}}, \frac{7}{30 N_\text{tst} n_p} \right) .
    \label{eq:rmse_theo_dist}
\end{equation}
This is the RMSE average and variance for a completely random model while, a trained one should return lower values. The full derivation is given in Appendix~\ref{app:dataset_size}.

\subsection{Model details}

\begin{figure}
    \centering
    \includegraphics[scale=0.75]{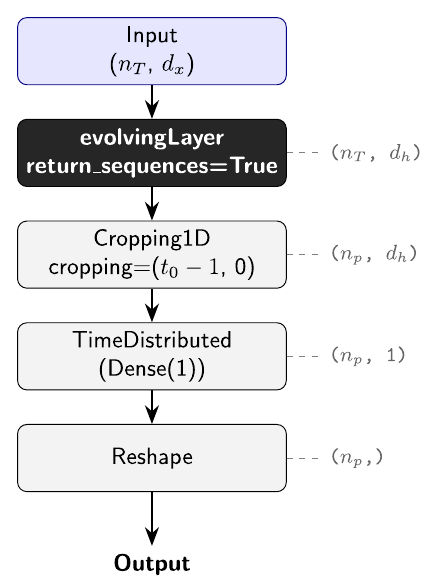}
    \caption{Diagram of CRNN sequential model in TensorFlow \cite{tensorflow2015-whitepaper} used to build the classical models. \texttt{evolvingLayer} must be a \texttt{keras.src.layers.rnn}-type object.}
    \label{fig:rnn_diagram}
\end{figure}

We have defined the CRNN model family with a sequential architecture detailed in Fig.~\ref{fig:rnn_diagram}. The black box is an evolving layer which can be SimpleRNN, LSTM or GRU.

\begin{figure*}
    \centering
    \includegraphics[scale=1]{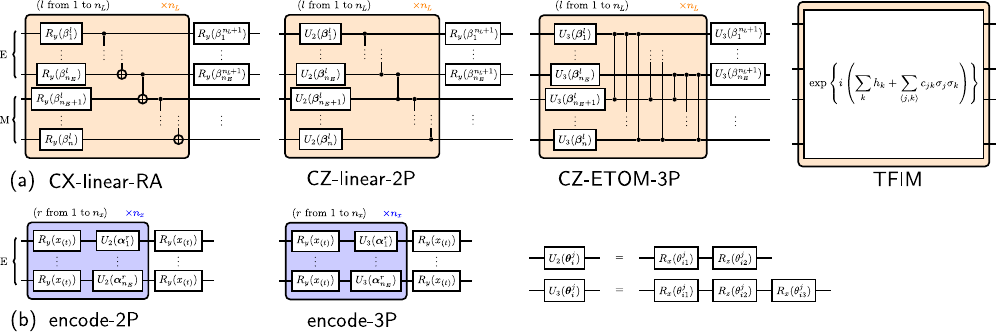}
    \caption{Collection of ansatze used in this benchmark. (a) Evolution operators. TFIM is the time evolution of a Transverse Field Ising Hamiltonian with tunable field and couplings. The evolution time is absorbed in the coefficients, which are trainable parameters. (b) Encoding operators.}
    \label{fig:qrnn_pieces}
\end{figure*}

The quantum models considered are built from the components summarized in Fig.~\ref{fig:qrnn_pieces} plus an output layer. The output layers are specific forms of Eq.~\ref{eq:output_layer_general},
\begin{equation}
    \begin{array}{ll}
        \text{exp-Z:} & \gamma_0 + \braket{Z} \\
        \text{full-base:} & \gamma_0 + \gamma_1 \braket{X} + \gamma_2 \braket{Y} + \gamma_3 \braket{Z}
    \end{array} ,
\label{eq:outlayer_component_definition}
\end{equation}
where $\gamma_0$ is a trainable bias. The selection of these architectures responds to the need to compare a broad range of circuit components. We found that three hardware-efficient ansatze (HEAs) with different properties and a reservoir-like model should serve as a meaningful group for establishing conclusions. On the one hand, HEA have been widely used in VQA studies, as a NISQ-friendly alternative for gate-based quantum computers, although they require a considerable number of trainable parameters. On the other hand, reservoir-like architectures are more suited for analog quantum computers, while reducing the number of trainable parameters. With the reservoir-like model, Hybrid, we indeed intend to see what happens when we reduce the number of circuit parameters, while measuring more information about the quantum states through a base of observables.

Using these components, we can build the architectures in Tab.~\ref{tab:quantum_models} for benchmark and comparison. The three HEA models gradually increase their complexity in terms of rotation gates. EMCX1 is intended to have only real amplitudes and thus it can only explore a part of the Hilbert space, which is the set of states without complex amplitudes. This setting substantially reduces the number of trainable parameters, but it is restricted. Meanwhile, EMCZ3 contains $U_3$ local gates that can perform any local rotation around an arbitrary axis in the Bloch sphere, and full entanglement between qubits of both registers E and M, at the cost of a large number of parameters. Finally, EMCZ2 represents a compromise between these two architectures: it retains substantial expressivity while requiring fewer parameters than EMCZ3. The Hybrid model resembles a quantum reservoir, by implementing controlled evolutions of a quantum dynamical system, while recovering all the degrees of freedom of the quantum state at measurement.

Special mention should be made of the encoding operators. We introduce the data re-uploading technique, that consists of introducing the input data multiple times in rotation gates, alternated by $n_x$ parameterized rotation gates to enhance the expressivity of quantum circuits \cite{perez-salinas_data_2020,schuld_effect_2021}. It can be critical, because even in the simplest task, such as reproducing an identity function, it may not reproduce exactly that function. This happens because it generates a Fourier series on inputs. Prior to starting the training of models, we studied what is the lowest data re-uploading number $n_x$ finding the equilibrium between excellent metric $R^2$ and low RMSE cost, as well as the normalization factor just before data is encoded, $c_\text{norm}$. This normalization factor must be low enough to avoid periodicity within the data interval, but large enough to capture non-linearities and to be sufficiently far from the numerical precision limit imposed by a hypothetical number of shots if these algorithms were deployed on quantum hardware. We analytically expand the Fourier series of expectation values measured from a quantum circuit made up of a single encoding operator with data re-uploading and find the minimum number of terms. The procedure is detailed at App.~\ref{app:re_uploading}. For consistency among models with complex amplitudes (all quantum except EMCX1), we set a common $n_x$, while for EMCX1 we set $n_x=0$ because any $n_x \geq 1$ would introduce complex amplitudes.

The sets of hyperparameters are summarized in Tab.~\ref{tab:hyperparameters}. We need to note several aspects of this choice of hyperparameters. First of all, the range of study for the memory size is up to 6, because sets of simulations beyond this limit start to be unaffordable due to the simulation time. Second, the number of exchange qubits, $n_E$, is always set to 1, because we consider encoding strategies where each input component is encoded in a qubit and the datasets are univariate time series. If the series were multivariate, using the encoding operators in Fig.~\ref{fig:qrnn_pieces} (b), we would have as many qubits in register E as input variables. Third, with HEA models, we choose the number of layers $n_L$ to equal the total number of qubits, $n_E+n_M$. This choice is made because, with the used ansatze, the light cone spans to all qubits when the number of layers $n_L$ equals the
total number of qubits.
The light cone determines which qubits and gates causally influence the expectation value of a local observable \cite{braccia_computing_2024,leone_practical_2024}. With this configuration, we make sure that a measurement at a single timestep can be influenced by the last memory qubit if needed. Fourth, we set the data re-uploading number $n_x$ after the analytical procedure was performed. Numerical results are detailed in Sec.~\ref{sec:results}. Fifth, we set the basis observables to $\{X,Y,Z\}$ in the Hybrid model, which is the full basis set of Pauli observables for 1 qubit.

\begin{table*}[]
    \small
    \centering
    \begin{tabular}{clll}
        \textbf{Name}     & \textbf{Components}                       & \textbf{Description}               \\
        EMCX1   & encode-2P + CX-linear-RA + exp-Z         & HEA with real amplitudes \\
        EMCZ2 & encode-2P + CZ-linear-2P + exp-Z          & HEA with reduced parameters \\
        EMCZ3   & encode-3P + CZ-ETOM-3P + exp-Z            & HEA with redundant parameters and high entanglement \\
        Hybrid  & encode-2P + TFIM + full-base          & Hybrid (digital + analog) with reservoir-like output layer
    \end{tabular}
    \caption{Families of quantum models used in the benchmark, where the components are defined in Fig.~\ref{fig:qrnn_pieces} and Eq.~\ref{eq:outlayer_component_definition}.}
    \label{tab:quantum_models}
\end{table*}

\begin{table*}[]
    \small
    \centering
    \begin{tabular}{lll}
         \textbf{Model}     & \textbf{Fixed} & \textbf{Variable} \\
         EMCX1 & $n_x=0$, $n_E=1$, $c_\text{norm} = \pi / 2$ & $n_M = \{ 1,2,3,4,5,6\}$, $n_L = n_M+1$ \\
         EMCZ2 & $n_x=3$, $n_E=1$, $c_\text{norm} = \pi / 2$ & $n_M = \{ 1,2,3,4,5,6\}$, $n_L = n_M+1$ \\
         EMCZ3 & $n_x=3$, $n_E=1$, $c_\text{norm} = \pi / 2$ & $n_M = \{ 1,2,3,4,5,6\}$, $n_L = n_M+1$ \\
         Hybrid & $n_x=3$, $n_E=1$, $c_\text{norm} = \pi / 2$ & $n_M = \{ 1,2,3,4,5,6\}$
    \end{tabular}
    \caption{Hyperparameter sets compared in this study.}
    \label{tab:hyperparameters}
\end{table*}

\subsection{Training details}

Both QRNN and CRNN models are trained by minimizing the per-sample mean squared error (MSE) over the prediction window of $n_p$ timesteps. Optimization is performed using Adam \cite{kingma_adam_2017}. It is an adaptive optimizer, with dynamic learning rate per trainable parameter. This is a property that makes it very suitable for the tasks of interest because it accelerates convergence and stabilizes the training, avoiding the use of natural gradient optimizers, which require substantially more quantum circuit evaluations. The learning rate was set to 0.001 for all trainings, because it is the standard in TensorFlow.

Each gradient evaluation is computed with a single sample, i.e., batch size is 1. Gradients are computed through JAX automatic differentiation, evaluated on simulated ideal circuits \cite{bradbury_jax_2018}. For the sake of simulation efficiency, neither shot noise nor hardware noise were considered across this study. Previous studies show that the effect of shot noise is not meaningful for large shot budgets \cite{viqueira_density_2025}. However, hardware noise may have either negative effects, such as inducing barren plateaus, or beneficial effects, such as helping optimization escape saddle points\cite{wang_noise-induced_2021,singkanipa_beyond_2025,liu_stochastic_2025}.
The cost function is RMSE, which is monitored across the validation set at the end of an epoch. It is computed by
\begin{equation}
    \text{RMSE}_\text{val} = \sqrt{\frac{1}{N_\text{val}}\sum_{s \in \mathcal{S}_\text{val}} \text{MSE}_s} ,
\end{equation}
where $s$ is the sample index and $N_\text{val}$ is the number of validation samples.

The model is considered to be trained at the lowest validation cost after $N_\text{ep} = 2000$. After all the experiments are finished, we check that this setting ensures training convergence in most cases. We consider that convergence was achieved when the moving average of both $C_\text{tr}$ and $C_\text{val}$ does not decrease below 0.01 in the last $N_C$ epochs. A moving average window of 50 epochs is applied because we need to remove the oscillations due to sample shuffling. The 0.01 threshold is chosen because it is an order of magnitude below the significant digits of the used cost function, RMSE.

The initialization of parameters at the beginning of the training is uniformly random for all models and tasks, for consistency. However, the interval of random generation changes. Classical models are initialized in the [-1,1] interval, which leads to a completely uninformed initial state. 
HEA quantum models are initialized to $[0,2 \pi]$ because we want to start in rotation-gate angles over a broad interval, that have a period of $2 \pi$. Finally, the Hybrid model initializes its parameters in the [0,1] interval. In this case, it is chosen as an arbitrary normalization of initial parameters, because we do not know the energy spectrum of the Hamiltonian describing the evolution. In general, the analog evolution is not periodic in the couplings and we would need to diagonalize the Hamiltonian to see its spectrum, which iteratively changes as parameters (transverse field and couplings) are changed. Thus, we cannot determine the parameter initialization by an objective criterion as done in HEA-based models.

\subsection{Gathering results and quantifying capacity}

We define the capacity of a model for solving a given task as the cutoff $\tau_c$. Specifically, we measure memory capacity and nonlinear capacity by evaluating performance on the STM and NARMA tasks using the square of the Pearson correlation coefficient $R^2$ as the metric, also called ``capacity'' in earlier QRC studies \cite{mujal_time-series_2023}. This metric is computed on the test set; therefore, it corresponds to $M_\text{tst}$.

We choose a threshold of $\varphi_c=0.9$, which marks an excellent level of performance in forecasting tasks in highly physical or strongly trending time series. Moreover, $0.9 \leq R^2 \leq 1$ is the approximate interval where the models could not fit perfectly due to imperfect encoding or statistical deviations due to data. We consider $\mu(\tau)$ a function that drastically changes its behavior between two regimes: a saturation regime, where the maximization metrics stay stable at a maximum value before the decay regime. Therefore, the selected $\tau_c$ should capture the full saturation regime and the initial part of the decay.

The probability distributions of $R^2$ per $\tau$ can be approximated by beta distributions. This is a simplification of the experimental distributions, which may be more accurately described by a mixture of beta distributions. However, simple beta distributions are enough for our purpose for two reasons. First, the values of $R^2$ are constrained in an interval [0,1], like the beta distribution (assuming that it is defined in the boundaries and its value is 0). Second, the shape is not as important as the cumulative distribution function in the 95 \% central interval, and this can be accurately captured through the beta distributions.
Because only ten observations are available for each value of $\tau$, this sample size is insufficient to estimate each distribution reliably in isolation. However, we assume that the distributions change smoothly with $\tau$. Thus, a good option is to perform a parametric fit of these beta distributions in terms of $\tau$. To this end, the parameters $\mu$ (mean) and $\phi$ (precision) are modeled as functions with a prior of smoothness through a generalized additive model (GAM) for location, scale and shape \cite{yee_vector_1996,rigby_generalized_2005,wood_smoothing_2016,ferrari_beta_2004}.

The method consists of modeling the effect of the covariate $\tau$ on both parameters using penalized cubic B-splines. The additive predictors are expressed as $\eta_\mu = \textbf{B} w_\mu$ and $\eta_\phi = \textbf{B} w_\phi$, where $\textbf{B}$ is the B-spline design matrix and $w$ represents the concatenated target weight vectors. This dual-spline formulation enables the model to dynamically capture variations in both the expected value and the underlying data dispersion across the covariate space without imposing rigid parametric shapes. The functions mapping beta parameters $\{\mu,\phi\}$ to additive predictors $\eta_{\{ \mu,\phi \}}$ are logit and logarithmic link functions, respectively, while the inverse functions are
\begin{equation}
    \begin{array}{ll}
        \mu & = \frac{1}{1+ \exp (- \eta_\mu)} \\
        \phi & = \exp (\eta_\phi)
    \end{array}
    \label{eq:prior_functions}
\end{equation}
which enable only the valid values of the parameters of the beta distribution: $\mu \in (0,1)$ and $\phi > 0$.
The fit of the weights is performed through maximum likelihood estimation (MLE) with the L-BFGS-B optimizer \cite{zhu_algorithm_1997}. The roughness penalty avoids abrupt changes in consecutive weights and the coefficients are set to $\lambda_{\{\mu,\phi\}} = 0.1$.

After fitting the distributions, we determine the 95 \% central interval of each $\mu$ through parametric bootstrap with 300 repetitions. Then, we compute the percentiles 2.5 \% and 97.5 \% for each modeled beta distribution. Finally, we compute the empirical coverage of the real points with respect to the modeled distributions as the portion of observations included within the 95 \% central interval, which must be around 95 \%.

\subsection{Evaluating QMI on trained models}

After evaluating the metrics, we test the QMI plateau in all models, after training, considering the trained model at the end of the epoch of minimum $C_\text{val}$. We check that the transitory phase estimated with random recurrent circuits is an appropriate bound for all models. We start from the premise that the circuit state is stationary when the QMI shows no trend after some initial period. In individual samples, the QMI fluctuates across timesteps but typically shows a trend only at the beginning. On average, we observe behaviors similar to Fig.~\ref{fig:max_washout}. In our experiments, $n_T=30$, and we define three hypothetical phases: (i) transitory ($1 \leq t \leq 10$), (ii) stationary ($11 \leq t \leq 30$), and (iii) full ($1 \leq t \leq 30$).

To rigorously assess whether the stationary phase exists, we employ a Bayesian evaluation of a weighted least squares (WLS) regression computed over the local data windows. The WLS regression assigns each observation a weight inversely proportional to its variance ($1/\sigma_i^2$) to neutralize the influence of outliers, yielding a robust slope estimate and its associated standard error. We establish a region of practical equivalence (ROPE), an interval $[-0.01,0.01]$, representing a magnitude physically indistinguishable from zero, taking into account that distinguishable QMI values are around an order of magnitude above. Assuming an uninformative prior, the posterior probability of the true slope is modeled analytically as a Student's t-distribution parameterized by the WLS slope estimate, the standard error, and the appropriate degrees of freedom, $N-2$, being $N$ the number of data points to fit and 2 the number of parameters in a linear regression, intercept and slope. By integrating this posterior distribution strictly within the ROPE boundaries, the test computes the explicit Bayesian probability that the underlying trend within a window is practically flat. If this probability is high we can assert that a window is effectively stationary, while if it is too low, we can consider that a window corresponds to a transitory phase.

\section{Results \label{sec:results}}

\begin{figure}
    \centering
    \includegraphics[scale=1]{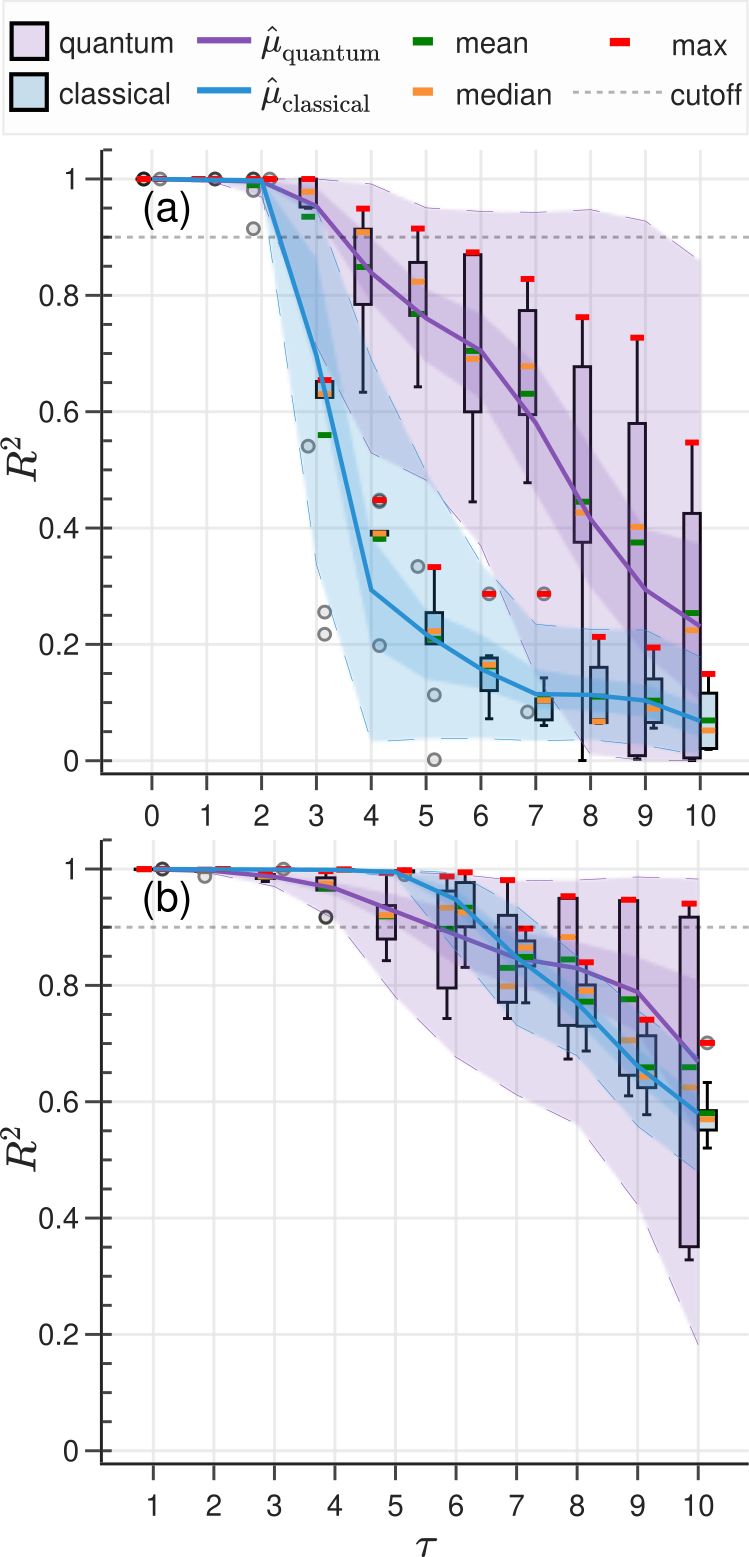} 
    \caption{Boxplots and GAM fits of training metrics in test sets after 10 training runs per $\tau$ for a quantum and a classical model. Outer shaded regions delimited by shaded lines represent the 2.5 \% and 97.5 \% percentiles of the distributions estimated by the GAM fit. Inner shaded regions represent the uncertainty of the fit of $\mu$. (a) STM-task --quantum model: EMCZ2 ($n_E=1$, $n_M=3$, $n_L=4$); classical model: VanillaRNN ($\text{units}=3$). (b) NARMA-task --quantum model: EMCZ2 ($n_E=1$, $n_M=6$, $n_L=7$); classical model: LSTM ($\text{units}=6$).}
    \label{fig:boxplots_capacity}
\end{figure}

After completing the STM, NARMA and MG tasks with seven model architectures and six hyperparameter configurations, we obtained the distributions of $R^2$ versus $\tau$. The median and mean $R^2$ systematically decrease with $\tau$ for both classical and quantum models. Representative examples are shown in Fig.~\ref{fig:boxplots_capacity}, where we compare classical and quantum models trained on STM and NARMA datasets.

Based on these results, we can highlight different aspects of the post-training metric distributions that are systematic across all tasks and models.

First, there is a plateau where both models reach the highest score with minimal variance. In theory, in the STM task the memory vectors are long enough to store past inputs independently, which is the optimal solution. These results show that, even when starting from a minimal-information state before training (i.e., uniform random initialization of parameters), gradients can guide optimization to a valid solution. This behavior is also observed in the other tasks under reasonable approximations.

These values of $R^2$ cannot be achieved with quantum models if the encoding strategy is not expressive enough. Thus, we conduct the analytical procedure to compute the minimum acceptable value $n_x$ and $c_\text{norm}$. We conduct experiments with input normalization $c_\text{norm} = \pi / 2$ and a data re-uploading number of $n_x=3$, which ensure both cost function (RMSE) and metric ($R^2$) quality. Specifically, with this configuration and the previous data normalization, we have feature maps giving
\begin{equation*}
    \begin{array}{lll}
        \text{(cost)}   & \text{(RMSE)}     & 0.0521 \\
        \text{(metric)} & \text{(} R^2 \text{)} & 0.9722
    \end{array} .
\end{equation*}
These quantities are tabulated in Tab.~\ref{tab:costs_fourier2} App.~\ref{app:re_uploading} and they would be the extremal values that could be reached in the limit of infinite random data samples if the quantum feature map were simply the encode-2P operator and $\langle Z \rangle^{\otimes n_E}$ expectation value. The scores we reach with QRNN are better because it consists of an encoding plus an evolving operator. However, if $n_x$ is not high enough, the $R^2$ score is known to be significantly lower than 1.

Second, once the plateau behavior is exceeded, the mean $R^2$ decreases and the performance distributions become more dispersed. In case (a), the CRNN median $R^2$ drops abruptly, whereas the QRNN median decreases more gradually. In case (b), the CRNN performs better before the cutoff but then decays faster than the QRNN. These distributions allow us to evaluate not only central tendencies (e.g., the mean) but also dispersion, which generally increases more in QRNNs.

\begin{figure*}
    \centering
    \includegraphics[scale=1]{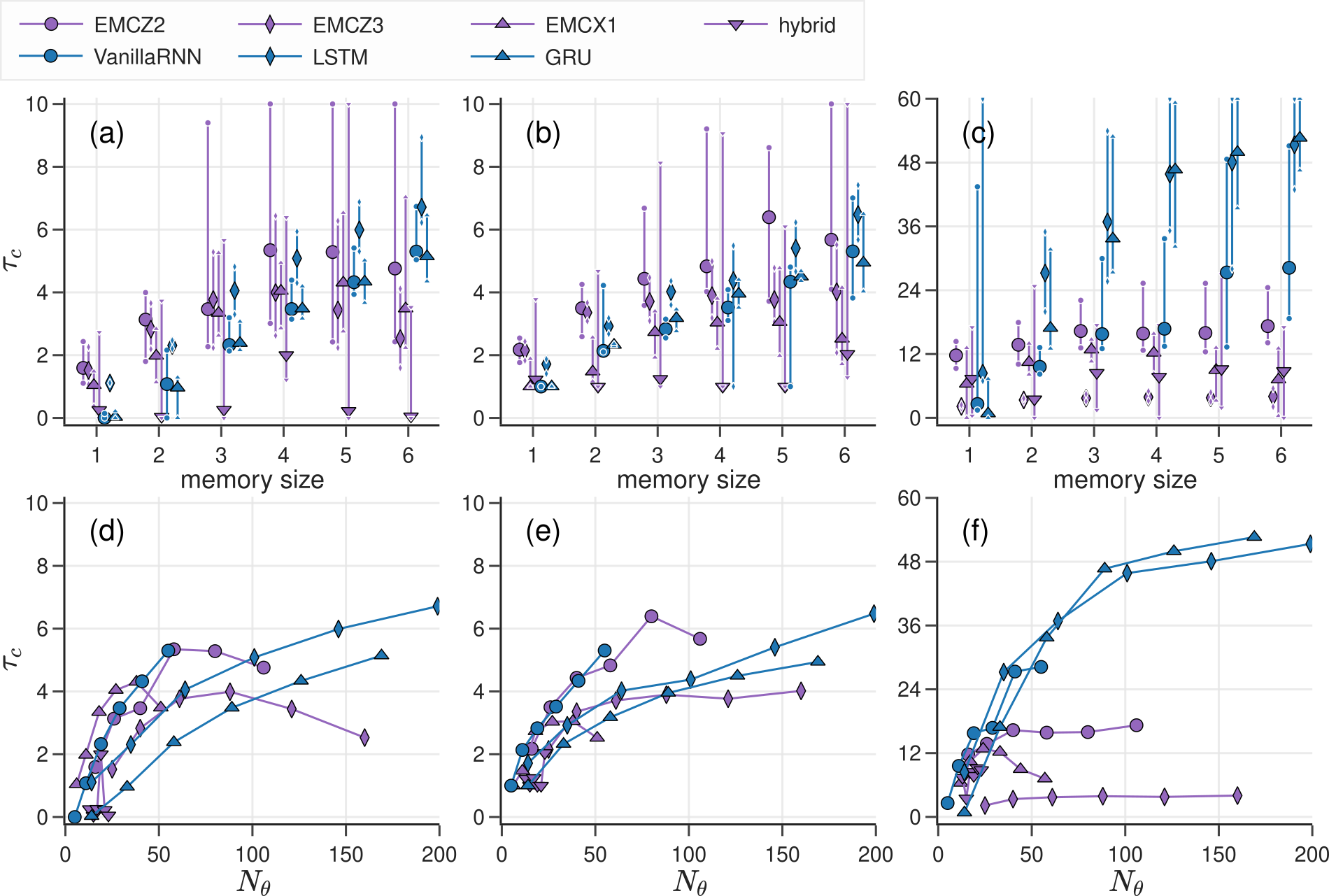}
    \caption{Summaries of the cutoffs at $R^2=0.9$ versus the memory size for all the models. (a,d) STM task, (b,e) NARMA task, (c,f) Mackey--Glass. (a,b,c) show the interpolation of $\mu$ at a threshold $R^2=0.9$ as large markers with black edges and the interpolation of minimum and maximum percentiles of the estimated beta distribution as small markers with white edges. Vertical lines thus provide the central 95 \% probability interval at the threshold, versus memory size. The empirical coverage of GAM fits is never below 90 \%. (d,e,f) show the interpolation of $\mu$ versus the number of trainable parameters.}
    \label{fig:condensed_cutoffs}
\end{figure*}

In Fig.~\ref{fig:condensed_cutoffs} we show the capacities $\tau_c$ at $\varphi_c = 0.9$ for all the models studied and the three tasks against memory size. Quantum models always struggle with larger memory size, except for EMCZ2 on the NARMA task, where we see a capacity advantage of the quantum model. Our results suggest that, for a constant number of epochs, the given training configuration and random parameter initialization, the increase of memory size in quantum models does not provide, in general, an advantage with respect to classical ones on average. However, we see that a good design can lead to better capacities at high memory size for a given task.

Our results show that higher capacity can be achieved with some quantum models in some tasks, as indicated by the upper tails of the distributions. This is particularly relevant because some trained quantum models, starting from an uninformed random initialization of parameters, converge to better solutions than classical ones. Therefore, a key challenge is finding a good initialization strategy. As a drawback, our results also suggest that, as the memory (system) size grows, finding a suitable initialization becomes more difficult.

Additionally, we report $R^2$ as a function of the number of trainable parameters (weights in classical neural networks). The scaling is similar for classical models and quantum HEA designs: the dominant contribution is quadratic in memory size. This occurs because we add as many layers as qubits in the quantum models, while classical models require $N\times N$ matrices to map one hidden vector to another. Therefore, the $R^2$-versus-memory trend is preserved when plotting against the number of parameters, $N_\theta$. The Hybrid architecture, which is designed to reduce the number of parameters, systematically achieves worse average scores, but it requires substantially fewer trainable parameters, scaling linearly with memory size.

In terms of comparative performance, as shown in Fig.~\ref{fig:condensed_cutoffs}, no single architecture performs best across all tasks. It always depends on the task to solve. HEA architectures show comparable performance at small memory sizes in STM and NARMA tasks, but EMCZ3 performs worse in MG. Meanwhile, Hybrid models are never better on average in terms of memory size, but they are comparable in terms of $N_\theta$.

If we focus on HEA models, EMCZ3 achieves peak performance at memory size 4, which is a relatively small value. Likely due to its overparameterization (groups of three parametric rotation gates) and the high entanglement generated by the E-to-M connections, its scaling with memory size appears poor across tasks. The EMCX1 model lacks complex amplitudes in its operators and state description and can therefore be simulated with half the state-vector storage, assuming identical precision and storage of real versus complex amplitudes. Within the HEA family, it requires the fewest parameters; however, as with Hybrid, it may not be expressive enough for these tasks, and its average performance starts to decrease around memory size 4 or 5. Finally, EMCZ2 strikes a balance between parameter count and performance, leading to the best average performance overall. It not only performs better on average, but in NARMA it also extends its range of good performance substantially beyond the other HEA models and the classical baselines. This suggests that quantum models can achieve higher capacity. However, this conclusion depends on good initialization: if good initializations become harder to find as memory size increases, the advantage could vanish.

Focusing on Hybrid, a plausible explanation for its poorer performance on average is that this architecture does not have enough parameters to reach the degrees of freedom required to fit the function generating the time series. Although it resembles QRC architectures, it lacks a high-dimensional vector of expectation values entering the output layer because it is limited to a single exchange qubit. Moreover, iterative training of the Hamiltonian that generates the evolution is challenging because it can induce complex energy spectra. Despite this average behavior, the cutoff intervals reach the extreme values in some cases, indicating that this architecture is particularly sensitive to training initialization. If we could find an adequate initialization strategy, this architecture could achieve markedly higher performance.

Finally, we compare the quantum architectures studied against the given classical baselines. HEA architectures perform similarly to classical architectures in STM and NARMA tasks, especially when the memory sizes are low. However, as stated before, EMCX1 and EMCZ3 perform worse on average for memory sizes greater than 4. The only quantum architecture that is competitive on average in the full range of study is the EMCZ2. Despite its apparent final drop in NARMA, this is inconclusive. In general, no quantum architecture performs better than the classical baselines. While VanillaRNN shows comparable performance for small memory sizes, it surpasses quantum models for large memory sizes, despite being a simple model. Meanwhile, LSTM and GRU surpass quantum models from memory size 2 onward in MG task, likely due to their capacity of selecting short- and long-term dependencies with their sophisticated architecture. A possible improvement would be to develop more sophisticated QRNN architectures.

It is important to note that within the range of study of $\tau$, in some cases it is not possible to see the complete span of the cutoff. Additional training runs with larger values of $\tau$ would be needed to determine these limits. However, from the initial design, we set the limit to 10 in NARMA and STM and 60 in MG in order to bound the number of trainings, which require expensive loops.

Regarding parameter efficiency, our results indicate that EMCZ2 --previously studied in \cite{viqueira_density_2025}-- provides the best balance, while showing comparable or better performance than classical baselines in the STM and NARMA tasks. EMCZ3 is the least efficient quantum architecture, as it requires almost as many parameters as the more sophisticated LSTM and GRU models while delivering worse performance. Both EMCX1 and Hybrid use fewer parameters, but they are not competitive with the baselines.

Finally we checked the convergence in terms of the training and validation cost curves, as described in Sec.~\ref{sec:methods}, as well as the phases of the quantum recurrent models through time in terms of the QMI.

We checked that, taking the last $N_C=500$ epochs, the moving-averaged cost curves decayed 0.01 or less in 95.6 \% of the quantum model trainings, while for $N_C=100$ this happened in 99.5 \%. Meanwhile, in classical models it happens in fewer cases: 86.6 \% for $N_C=500$ and 98.8 \% for $N_C=100$, respectively. Thus, under the training protocol considered here, 2000 epochs were enough for convergence in the vast majority of cases. In any case, as the results indicate, this convergence does not guarantee the best possible solution, which depends strongly on parameter initialization.

We also checked that in 99.3 \% of the trainings the posterior probability that the slope of QMI lies within the ROPE is over 66 \% in the stationary phase, while this only happened for the 6.8 \% in the transitory phase. The ROPE probability in the transitory phase is greater than or equal to that in the stationary phase in only 2.6 \% of cases, while in only 0.3 \% it is strictly greater. Therefore, in a vast majority of cases, we can be confident that in timesteps from 11 to 30, the QRNN is in a stationary phase, while from 1 to 10, it works in a different regime. The choice of a 10-timestep transitory phase seems reasonable in consequence. Even in cases where we are not confident about the stationary condition, the system has additional time to leave the transitory regime in the context window (11-to-20 timesteps). For more details about this analysis and examples, we refer the reader to App.~\ref{app:mutual_info}.

\section{Conclusions \label{sec:conclusions}}

In this work, we introduced and applied a methodology to quantify the performance of quantum recurrent models across STM, NARMA, and Mackey--Glass tasks in comparison with well-known classical baseline models: the VanillaRNN, and sophisticated models LSTM and GRU. This enabled a rigorous analysis of memory, nonlinear, and prediction capacity in relation to memory size. In addition, we measured the quantum mutual information between the QRNN memory and exchange registers, which allowed us to determine the length of a transitory phase during which the system decays from its initial conditions before making predictions.

To make a global comparison, rather than analyzing raw performance metrics, we analyzed how performance scales with memory size. Our results show no advantage for any quantum model in the STM task. Although some quantum models show slightly better results (inconclusive given the uncertainty bounds) at small memory sizes and small numbers of trainable parameters, average performance stops improving—or even degrades—at larger sizes. In contrast, within the studied ranges, the classical networks continue to improve. The NARMA task yields more promising results for the EMCZ2 quantum model, which shows better average performance across the entire range studied under different random parameter initializations. Moreover, the distributions exhibit higher upper tails for the EMCZ2 and Hybrid models, depending on the task, indicating that performance can depend on parameter initialization as much as on architecture. In the Mackey--Glass prediction task we conclude that, while at small memory size, performances are comparable on average across all models, classical models are significantly better at higher memory sizes.

Several directions remain for future work. First, because gradient-descent training is sensitive to parameter initialization, developing effective initialization strategies for QRNNs is a priority. Although established strategies, such as Glorot initialization, are available for classical neural networks \cite{glorot_understanding_2010}, no generally accepted data-agnostic initialization strategy has been established for quantum machine learning models. Accordingly, we used random initialization for all models in this work to ensure a fair comparison. Future research should therefore investigate QRNN-specific initialization schemes, as well as alternative optimization algorithms. Second, the scope of this study is limited primarily by the computational cost of simulating quantum networks, which grows exponentially with network and memory size at large system sizes. Deploying these models on quantum hardware offers a route to assessing their performance at larger scales, including the potential degradation caused by hardware errors and sampling noise, as well as the impact of communication latencies in CPU--QPU interactions on total execution times \cite{barral_review_2025}. Third, future work should extend the proposed architectures and benchmarking methodology to multivariate time series, thereby broadening the range of sequential-data problems to which they can be applied. Finally, given continuing advances in CPUs and GPUs, these models may also provide a basis for quantum-inspired algorithms running on classical hardware \cite{huynh_quantum-inspired_2027}. Taken together, these directions build on the initial steps presented here and provide a path toward the development, large-scale evaluation, and practical assessment of variational quantum recurrent neural networks.

\section*{CRediT authorship contribution statement}
\textbf{J.D. Viqueira}: Conceptualization, Methodology, Software, Formal analysis, Writing – review \& editing. \textbf{A. Gómez}: Conceptualization, Formal analysis, Writing – review \& editing. \textbf{D. Mera}: Formal analysis, Writing – review \& editing.

\section*{Declaration of interests}
The authors declare that they have no known competing financial interests or personal relationships that could have appeared to influence the work reported in this paper.

\section*{Acknowledgments}
We thank the CESGA Quantum Computing researchers for their feedback and the stimulating intellectual environment they provide. J.D.V. acknowledges Francisco Yáñez for their discussions about time series.

This work was supported by MICIU/AEI/10.13039/501100011033 and by ERDF/EU (grant 
PID2024-159713OB-I00). This work was partially supported by the Galician Government under grant ED431B 2024/44.
J. D. Viqueira was supported by Axencia Galega de Innovación (Xunta de Galicia) through the ``Programa de axudas á etapa predoutoral'' (grant IN606A-2023/011).

The simulations reported in this work were performed using Galicia Supercomputing Center (CESGA) FinisTerrae III supercomputer with financing from the Programa Operativo Plurirregional de España 2014-2020 of ERDF, ICTS-2019-02-CESGA-3, and the Qmio quantum infrastructure, with financing from the European Union, through the Programa Operativo Galicia 2014-2020 of ERDF$\_$REACT EU, as part of the European Union’s response to the COVID-19 pandemic.

\section*{Declaration of generative AI and AI-assisted technologies in the manuscript preparation process}

During the preparation of this work the authors used Google Gemini, Perplexity and Claude to develop code, search information  and review the language and written consistency of the manuscript. After using this tools/services, the authors reviewed and edited the content as needed and took full responsibility for the content of the published article.

\section*{Data availability statement \label{sec:code}}

The data that support the findings of this study are openly available at the following URL/DOI: \url{https://doi.org/10.5281/zenodo.24287060} (Datasets and results).

\bibliography{Paper_benchmark25}


\appendix 

\section{Determination of dataset sizes \label{app:dataset_size}}

We determine the size of the generated datasets to have reasonable levels of statistical deviations due to sample selection for training, validation and test partitions. A limited number of samples would lead to inaccurate evaluations of cost in validation and test. A way to escape from this is to perform cross-fold validation, but this would complicate the interpretability of our benchmarks and would require more computational resources. This is why we accomplish estimations of the minimum required sizes of datasets starting from simple statistical assumptions. Meanwhile, we take into account that the computational time scales approximately linearly with the number of samples, thus we need to keep moderate dataset sizes.

We choose the worst case scenario, which is uniformly random data and completely random (untrained) models. As models are trained, the statistical deviations are lower because the distances between predictions and targets are reduced. Our labels are independently drawn from a uniform distribution, $y_i \sim U(-1,1)$,
while a completely random model will also output $x_i \sim U(-1,1)$ ,
where for notational simplicity in this appendix we denote model outputs by $x$ instead of $\bar{y}$. 

Our loss function is
\begin{equation}
    \text{RMSE} = \sqrt{\frac{1}{N} \sum_{i=1}^N \left( x_i - y_i \right)^2} .
\end{equation}

Let's define the difference $d_i \equiv x_i - y_i$. It has a triangular distribution in the interval $[-2,2]$. As the expectation value operator is linear, the mean is
\begin{equation}
    E[d_i] = E[x_i] - E[y_i] = 0 - 0 = 0,
\end{equation}
while its variance is
\begin{equation}
\begin{split}
    \text{Var} (d_i) &= \text{Var} (x_i) + \text{Var} (y_i) - 2 \text{Cov} (x_i, y_i) \\
    &= 2 \times \frac{(1-(-1))^2}{12} = 2 \times \frac{4}{12} = \frac{2}{3} ,
\end{split}
\end{equation}
where we have used the well-known value for the variance of a rectangular distribution. $x_i$ and $y_i$ are independent, because the model is random and does not generate correlations.

Prior to computing the statistics for the RMSE, let's derive those of MSE.
\begin{equation}
    \text{MSE} = \frac{1}{N} \sum_{i=1}^N d_i^2.
\end{equation}

Defining $z_i \equiv d_i^2$, this is a random variable in the interval $[0,4]$. The MSE is actually the sample mean of the i.i.d. variable $z_i$.

Its mean is
\begin{equation}
    E[z_i] = E[d_i^2] = \text{Var}(d_i) + (E[d_i])^2 = \frac{2}{3} + 0^2 = \frac{2}{3},
\end{equation}
while its variance,
\begin{equation}
\begin{split}
        \text{Var} (z_i) =& E[z_i^2] - (E[z_i])^2 = E[d_i^4] - (E[d_i^2])^2 \\
        =& \frac{16}{15} - \frac{4}{9} = \frac{28}{45} .
\end{split}
\end{equation}

For large numbers of samples, $N$, different measurements of the MSE follow a normal distribution with expected value (average) $\mu$ and standard deviation $\sigma$, according to the central limit theorem. $\mu$ and $\sigma^2$ are the computed mean and variance divided by the number of samples, respectively.

\begin{equation}
    \text{MSE} \approx \mathcal{N} \left( \frac{2}{3}, \frac{\text{Var} (z_i)}{N} \right) = \mathcal{N} \left( \frac{2}{3}, \frac{28}{45 N} \right) .
\end{equation}

Finally, to compute the RMSE, we use the delta method.
The expectation value is then
\begin{equation}
    E[\text{RMSE}] = E[g(X)] \approx g(\mu_X) \approx \sqrt{\frac{2}{3}} ,
\end{equation}
and the variance,
\begin{equation}
\begin{split}
    \text{Var}(\text{RMSE}) &= \text{Var}(g(X)) \approx [g'(\mu_X)]^2 \text{Var}(X) \\
    &= \left[ \frac{1}{2} \frac{1}{\sqrt{2/3}} \right]^2 \frac{28}{45N} = \frac{7}{30N} .
\end{split}
\end{equation}
Therefore,
\begin{equation}
    \text{RMSE} \approx \mathcal{N} \left( \sqrt{\frac{2}{3}}, \frac{7}{30N} \right) ,
    \label{eq:rmse_theo_dist-app}
\end{equation}

In our benchmarks, we generate series of $N=3000$ timesteps, which are divided into disjoint windows of $n_T=30$, and just $n_p=10$ outputs are used for training and prediction. With a 70\%-15\%-15\% distribution for training, validation and test, the number of points per partition and the estimated standard deviation are detailed in Eq.~\ref{eq:normal_rmse}. 

\begin{equation}
    \begin{array}{lll}
        N_\text{tr} n_p =700 & \mu_\text{tr} = 0.816 & \sigma_\text{tr} = 0.018 \\
        N_\text{\{val,tst\}} n_p=150 & \mu_\text{\{val,tst\}} = 0.816 & \sigma_\text{\{val,tst\}} = 0.039
    \end{array}
\label{eq:normal_rmse}
\end{equation}

We also show the distributions, obtained numerically, and we check that the results are the same as with the mathematical derivations. For the smallest sample of 150 points (constructed from 15 test samples), there is at most 5 \% chance that the RMSE is beyond the $\mu \pm 2\sigma = 0.816 \pm 0.079$ interval because of the selection of finite training samples. That margin around the mean may be relevant in a quantity, the RMSE, that can acquire values between 0 and 2. In trained models, the RMSE will usually be close to and lower-bounded by 0, with small deviation.

\section{Quantum Mutual Information across timesteps \label{app:mutual_info}}

\begin{figure*}
    \centering
    \includegraphics[scale=1]{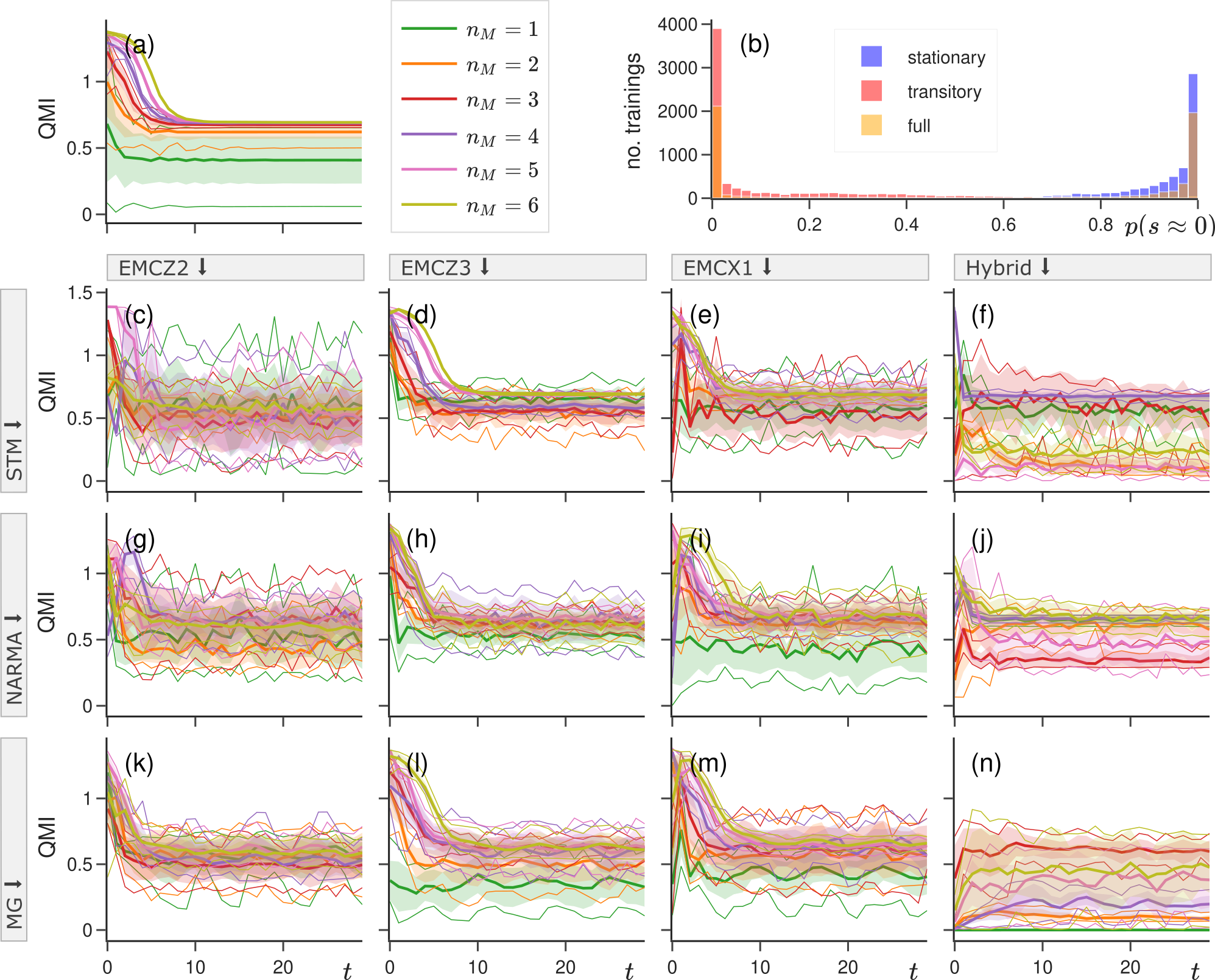}
    \caption{Details of QMI across timesteps for every QRNN configuration. (a) mean (thick solid line), standard deviation (shadow) and extreme values (thin solid line) across timesteps for 15 random networks with different seeds. (b) Summary of posterior probabilities that the slope lies within the ROPE in trained QRNN in the three partitions with test samples. Total of trainings is 7680; bin size is 0.02. (c-n) represent the same statistics as (a) in the test partition (15 samples) of a randomly chosen training per network configuration, evaluated at the best set of parameters found in the training. (c), (d), (e), (f) STM task; (g), (h), (i), (j) NARMA task; (k), (l), (m), (n) MG task. (c), (g), (k) EMCZ2; (d), (h), (l) EMCZ3; (e), (i), (m) EMCX1; (f), (j), (n) Hybrid.}
    \label{fig:qmi_details}
\end{figure*}

A good measure of the difference between the transitory and stationary phase is the QMI. In order to avoid making predictions in the transitory phase window, we estimate the minimum timestep from which the stationary phase starts for all the models. A QRRM with absence of structure is a good point to start. Of course, it is impossible to determine the real transitory phase before training the models. Thus, we start by assuming that this model will be fair enough to estimate the maximum length of the transitory phase. It consists of a QRNN with Haar random operators. This is supposed to produce maximum entropy. We tested sizes of $n_E=1$ and $n_M$ from 1 to 6, and the QMI of each configuration is detailed in Fig.~\ref{fig:qmi_details} (a).

We ran 150 experiments of QRRMs, and divided them in 15 groups of 10 samples, imitating the number of test samples in our trainings, and the number of experiments, $N_{exp}$ per hyperparameter set.

In Sec.~\ref{sec:methods} we describe how to determine whether a phase (group of timesteps) is stationary or not in a trained model. We show in Fig.~\ref{fig:qmi_details} (b) the distributions of the probabilities of slope to be in the interval [-0.01,0.01] across all the trainings together. We see that for most cases, the probability in the stationary is close to 1, while in the transitory phase is close to 0. Meanwhile, the full phase is more balanced.

\section{Data re-uploading selection \label{app:re_uploading}}

The selection of data re-uploading parameter, $n_x$, in EMCZ2, EMCZ3 and Hybrid models was motivated by the performance of STM-0 task, which is considered the baseline where the performance should be maximum, because it is the trivial map $f(x) = x$.

When normalized inputs in [-1,1] are normalized again to the interval $[-\pi,\pi]$, the analytical derivation of the MSE for the approximation of the quantum circuit to the target function is easy and we explicitly show it. That normalization applies only to inputs, while outputs are restricted to [-1,1]. Anyway, the results generalize directly by dividing the Fourier coefficients by $\pi$. In this case, we would have the map $f(x) = x / \pi$, $x \in [-\pi,\pi]$.

The repeated encoding of a value $x$ in a quantum circuit leads to a Fourier expansion \cite{schuld_effect_2021},
\begin{equation}
    S_F = \sum_{n=-F}^{F} c_n e^{inx} ,
\end{equation}
where $F = n_x + 1$, following the notation in the main text, i.e., $F=1$ means that a single rotation gate encodes the data. Given a target function $f(x)$ that applies in the interval $[-\pi,\pi]$, the coefficients leading to the closest Fourier approximation are
\begin{equation}
    c_n = \frac{1}{2\pi} \int_{-\pi}^{\pi} f(x) e^{-inx} dx .
\end{equation}

For $f(x) = x$, the coefficients are
\begin{equation}
\begin{array}{rl}
     c_0 &= 0  \\
     c_n &= i \frac{(-1)^n}{n} \text{ for } n \neq 0 . 
\end{array}
\end{equation}

Hence,
\begin{equation}
    S_F(x) =2\sum_{n=1}^{F}\frac{(-1)^{n+1}}{n}\sin(nx) \approx f(x) .
\end{equation}

The loss function is
\begin{equation}
    \text{MSE} = \frac{1}{2\pi} \int_{-\pi}^{\pi} |f(x) - S_F(x)|^2 dx .
\end{equation}

Using the Parseval theorem twice,
\begin{equation}
\begin{split}
    \frac{1}{2\pi} \int_{-\pi}^{\pi} | f(x) |^2 dx = \sum_{-\infty}^{\infty} | c_n |^2 = \sum_{n=-F}^{F} | c_n |^2 + \sum_{|n|>F} | c_n |^2 = \\ \frac{1}{2\pi} \int_{-\pi}^{\pi} |S_F(x)|^2 dx +  \frac{1}{2\pi} \int_{-\pi}^{\pi} |f(x) - S_F(x)|^2 dx .
\end{split}
\end{equation}
To interpret the final term,
\begin{equation}
    \sum_{|n|>F} | c_n |^2 = \sum_{-\infty}^{\infty} | c_n - \hat{c}_n |^2 ,
\end{equation}
where the second coefficients correspond to the truncated approximation $S_F$ and then,
\begin{equation}
    \left\{ 
    \begin{array}{cc}
        \hat{c}_n = c_n & \text{ if } |n| \leq F \\
        \hat{c}_n = 0   & \text{ if } |n| > F
    \end{array}
    \right. .
\end{equation}

Solving for the approximation error,
\begin{equation}
\begin{split}
    \frac{1}{2\pi} \int_{-\pi}^{\pi} |f(x) - S_F(x)|^2 dx = \\ \frac{1}{2\pi} \int_{-\pi}^{\pi} | f(x) |^2 dx - \sum_{n=-F}^{n=F} | c_n |^2 ,
\end{split}
\end{equation}

Finally,
\begin{equation}
    \int_{-\pi}^{\pi} | f(x) |^2 dx = \frac{2}{3} \pi^3 .
\end{equation}

In the selected interval, we need close to 1000 Fourier terms to be able to approximate the target function below 0.05 RMSE. This happens because the Fourier series approximates well in the major part but close to the limits it diverges. As soon as we reduce the interval to $[ -0.95 \pi, 0.95 \pi ]$, we get the values of MSE, RMSE and $R^2$ given in Tab.~\ref{tab:costs_fourier1}, computed numerically.

\begin{table}[]
    \centering
    \begin{tabular}{|r|r|r|r|}
        $F$ & \textbf{Min. MSE} &  \textbf{Min. RMSE} & \textbf{Max. $R^2$} \\
        1 & 0.0931 & 0.3051 & 0.6914 \\
        2 & 0.0442 & 0.2101 & 0.8537 \\
        3 & 0.0245 & 0.1565 & 0.9192 \\
        6 & 0.0066 & 0.0812 & 0.9789 \\
        20 & 0.0011 & 0.0325 & 0.9965
    \end{tabular}
    \caption{Loss, cost and metric functions of a Fourier approximator for input data normalized to the $[ -0.95 \pi, 0.95 \pi ]$ range and outputs to $[-0.95,0.95]$.}
    \label{tab:costs_fourier1}
\end{table}

We still need high-order Fourier approximation with this normalization. However, a normalization factor of $\pi / 2$ will give us satisfactory results. Apart from this normalization factor, the series are normalized to the interval [-0.95,0.95]. Thus, the map to be found is 
\begin{equation}
    f: [-0.475 \pi,0.475 \pi] \to [-0.95,0.95] .
\end{equation}

The numerical results are attached in Tab.~\ref{tab:costs_fourier2}. We see that $F=4$ is sufficient to obtain an RMSE $< 0.1$, which is the minimum required for having an error of an order of magnitude smaller than the range of data, while having $R^2 > 0.95$. This apparent discrepancy between $R^2$ and RMSE is because they are different measures, whereas RMSE measures prediction error, $R^2$ measures squared linear correlation, which may be severely affected by the oscillations generated in the Fourier series.

\begin{table}[]
    \centering
    \begin{tabular}{|r|r|r|r|}
         $F$ & \textbf{Min. MSE} &  \textbf{Min. RMSE} & \textbf{Max. $R^2$} \\
         \textbf{1} & 0.0283 & 0.1681 & 0.9884 \\
         2 & 0.0076 & 0.0869 & 0.8999 \\
         3 & 0.0055 & 0.0742 & 0.9283 \\
         \textbf{4} & \textbf{0.0027} & \textbf{0.0521} & \textbf{0.9722} \\
         5 & 0.0022 & 0.0467 & 0.9747 \\
         6 & 0.0014 & 0.0372 & 0.9820
    \end{tabular}
    \caption{Loss, cost and metric functions of a Fourier approximator for input data normalized to the $[-0.475 \pi,0.475 \pi]$ range and outputs to $[-0.95,0.95]$.}
    \label{tab:costs_fourier2}
\end{table}

\section{Details of results \label{app:results_details}}
We attach the details of the GAM fit in Tab.~\ref{tab:gam_params}.

To understand the given parameters, let's explain some details of the GAM fit. We first generate univariate B-spline bases for features. A new feature matrix $\bm{B}$ is generated consisting of $n_\text{splines} =$ (maximum knots)+(degree)-1 spline basis functions of the given polynomial degree. The number of knots divides the X-axis in several segments to draw independent single curves of the given degree. The maximum of knots means that, if the x-axis has fewer unique values, the number of knots equals this number.

The prior functions described in Eq.~\ref{eq:prior_functions} take as argument $\eta_i = \textbf{B} w_i, i \in \{ \mu, \phi \}$, while the negative penalized log-likelihood measures the distance between real point distributions and the generated distributions by the prior functions evaluated on $\eta_i$, while applying L2 regularization terms to avoid overfitting.

Finally, a parametric bootstrap of 300 repetitions is accomplished to estimate the uncertainty of the fit of $\mu$. In each bootstrap iteration, observations are simulated from the fitted beta distributions, the model is refitted, and the resulting mean curve is stored. The variability across the 300 fitted curves quantifies uncertainty.

\begin{table}[]
    \centering
    \begin{tabular}{lr}
        maximum knots & 8 \\
        degree & 3 \\
        $\lambda_\mu$ penalty & 0.1 \\
        $\lambda_\phi$ penalty & 0.1 \\
        bootstrap & 300
    \end{tabular}
    \caption{Parameters to fit GAM.}
    \label{tab:gam_params}
\end{table}

\section{Simulation details \label{app:simulation_details}}
Simulations of classical networks were performed with TensorFlow, while quantum simulations were performed with a purpose-built emulator developed by the authors available at \url{https://doi.org/10.5281/zenodo.12755195}, enhanced with JAX. Versions are specified in Tab.~\ref{tab:software}. The longest training run was observed for the Hybrid model with $n_M=6$, which took 4 h 10 min in the longest training with 32 CPUs. The used hardware, specified in Tab.~\ref{tab:hardware}, is part of a high-performance computing infrastructure.

\begin{table}
    \centering
    \small
    \begin{tabular}{lll}
        \textbf{Package} & \textbf{Version} & \textbf{Used for}\\
        TensorFlow     & 2.21.0 & CRNN training and inference \\
        JAX            & 0.9.0.1 & QRNN training and inference \\
        Optax          & 0.2.7 & QRNN training\\
        JiTCDDE        & 1.8.3 & Generation of MG dataset\\
        Scikit-learn   & 1.8.0 & Analysis of results, GAM fits\\
         
    \end{tabular}
    \caption{Package versions and purposes.}
    \label{tab:software}
\end{table}

\begin{table}
    \centering
    \small
    \begin{tabular}{|l|l|l|}
        \hline
        \textbf{Node} & CPU node & GPU node  \\
        \hline
                             & 2 $\times$ Intel Xeon   & 2 $\times$ Intel Xeon   \\
        \textbf{Processor} & Ice Lake 8352Y   & Ice Lake 8352Y  \\
                            & 32 cores 2.2 GHz & 32 cores 2.2 GHz \\
        \hline
        \textbf{Memory} & 256 GB RAM & 256 GB RAM \\
        \hline
        \textbf{GPU} & \textit{None} & 2 $\times$ Nvidia A100  \\
         \hline
    \end{tabular}
    \caption{Hardware specifications.}
    \label{tab:hardware}
\end{table}

\end{document}